\documentclass[a4paper]{article}
\usepackage{RR}
\RRNo{6530}
\usepackage{hyperref}
\usepackage{amsmath,amssymb,amsfonts}
\usepackage[cm]{aeguill}

\usepackage{epsfig}
\usepackage{algorithm}
\usepackage{algorithmic}
\newcommand{\Ee}{\mathbb E}
\newcommand{\Oo}{\mathbb O}
\newcommand{\Rr}{\mathbb R}

\newcommand{\refeq}[1]{(\ref{#1})}
\def\ind{{\hbox{\rm l\kern-.4em\hbox{\rm l}}}~}

\RRdate{Mai 2008}

\RRauthor{
Viet\_Dung Doan\thanks[sfn]{INRIA, OASIS} 
\and
Abhijeet Gaikwad\thanksref{sfn}
\and
Mireille Bossy\thanks{INRIA, TOSCA} 
\and
Fran\c{c}oise Baude\thanksref{sfn}
\and
Ian Stokes-Rees\thanks{Dept. Biological Chemistry \& Molecular Pharmacology, Harvard Medical School}
}
\authorhead{Doan, Gaikwad, Bossy, Baude \& Stokes-Rees}
\RRtitle{Algorithmes de Pricing parall\`eles pour des Options Bermudiennes/Am\'ericaines multidimensionnelles par une m\'ethode de Monte
Carlo} 
\RRetitle{Parallel Pricing Algorithms\\
for Multi--Dimensional Bermudan/American Options \\
using Monte Carlo methods} 
\titlehead{Parallel Pricing Algorithms for Multi--Dimensional Bermudan/American Options}
\RRresume{Dans ce papier, nous pr\'esentons deux algorithmes de type
Monte Carlo pour le pricing d'options Bermudiennes/Am\'ericaines
multidimensionnelles. La premiere approche repose sur le calcul de la
fronti\`ere d'exercice, tandis que la seconde repose sur la
classification des  valeurs d'exercice et  de continuation. Nous \'evaluons les performances des
algorithmes dans un environnement grille. Nous montrons l'efficacit\'e
des approches propos\'ees dans un environnement h\'et\'erog\`ene. Nous
identifions les  contraintes d'\'evolutivit\'e  dues \`a la structure
algorithmique. 
}
\RRabstract{In this paper we present two parallel Monte Carlo based
algorithms for pricing multi--dimensional Bermudan/American
options. First approach relies on computation of the optimal exercise
boundary while the second relies on classification of continuation and
exercise values. We also evaluate the performance of both the
algorithms in a desktop grid environment. We show the effectiveness of
the proposed approaches in a heterogeneous computing environment, and
identify scalability constraints due to the algorithmic structure.} 
\RRmotcle{options Bermudiennes/Am\'ericaines
multidimensionnelles, M\'ethodes de Monte Carlo parall\'eles, Grid computing.}
\RRkeyword{Multi--dimensional Bermudan/American option, Parallel Distributed Monte Carlo methods, Grid computing.}
\RRprojets{OASIS et TOSCA}
\RRtheme{\THCom \THNum} 

\URSophia 

\begin{document}
\makeRR   

\section{Introduction}
Options are derivative financial products which allow buying and
selling of risks related to future price variations. The option buyer
has the right (but not obligation) to purchase (for a call option) or
sell (for a put option) any asset in the future (at its exercise date)
at a fixed price. Estimates of the option price are based on the well
known arbitrage pricing theory: the option price is given by the
expected value of the option payoff at its exercise date. For example,
the price of a call option  is the expected value of the positive part
of the difference between the market value of the underlying asset and
the asset fixed price  at the exercise date. The main challenge in
this situation is modelling the future asset price and then estimating
the payoff expectation, which is typically done using statistical
Monte Carlo (MC) simulations and careful selection of the static and
dynamic parameters which describe the market and assets. 

Some of the widely used options include American option, where the
exercise date is variable, and its slight variation Bermudan/American
(BA) option, with the fairly discretized variable exercise
date. Pricing these options with a large number of underlying assets
is computationally intensive and requires several days of serial
computational time (i.e. on a single processor system). For instance,
Ibanez and Zapatero (2004) \cite{ibanez2004mcv} state that pricing the
option with five assets takes two days, which is not desirable in
modern time critical financial markets. Typical approaches for pricing
options include the binomial method \cite{cox1979ops} and MC
simulations \cite{glasserman2004mcm}. Since binomial methods are not
suitable for high dimensional options, MC simulations have become the
cornerstone for simulation of financial models in the industry. Such
simulations have several advantages, including ease of implementation
and applicability to multi--dimensional options. Although MC
simulations are popular due to their ``embarrassingly parallel"
nature, for simple simulations, allows an almost arbitrary degree of
near-perfect parallel speed-up, their applicability to pricing
American options is complex\cite{ibanez2004mcv},
\cite{broadie1997vao}, \cite{longstaff:vao}. Researchers have proposed
several approximation methods to improve the tractability of MC
simulations. Recent advances in parallel computing hardware such as
multi-core processors, clusters, compute ``clouds", and large scale
computational grids have also attracted the interest of the
computational finance community. In literature, there exist a few
parallel BA option pricing techniques. Examples include Huang (2005)
\cite{huang2005pap} or Thulasiram (2002) \cite{thulasiram2016pep}
which are based on the binomial lattice model. However, a very few
studies have focused on parallelizing MC methods for BA pricing
\cite{toke2006mcv}.  In this paper, we aim to parallelize two American
option pricing methods: the first approach proposed in Ibanez and
Zapatero (2004) \cite{ibanez2004mcv} (I\&Z) which computes the optimal
exercise boundary and the second proposed by Picazo (2002)
\cite{hickernell2002mca} (CMC) which uses the classification of
continuation values. These two methods in their sequential form are
similar to recursive programming so that at a given exercise
opportunity they trigger many small independent MC simulations to
compute the continuation values. The optimal strategy of an American
option is to compare the exercise value (intrinsic value) with the
continuation value (the expected cash flow from continuing the option
contract), then exercise if the exercise value is more valuable. In
the case of I\&Z Algorithm the continuation values are used to
parameterize the exercise boundary whereas CMC Algorithm classifies
them to provide a characterization of the optimal exercise
boundary. Later, both approaches compute the option price using MC
simulations based on the computed exercise boundaries. 

Our roadmap is to study both the algorithms to highlight their
potential for parallelization: for the different phases, our aim is to
identify where and how the computation could be split into independent
parallel tasks. We assume a master-worker grid programming model,
where the master node splits the computation in such tasks and assigns
them to a set of worker nodes. Later, the master also collects the
partial results produced by these workers. In particular, we
investigate parallel BA options pricing to significantly reduce the
pricing time by harnessing the computational power provided by the
computational grid.  

The paper is organized as follows. Sections \ref{sectionibanezzapatero} and \ref{sectionpicazo} focus on two pricing methods and are structured in a similar way: a brief introduction to present the method, sequential followed by parallel algorithm and performance evaluation concludes each section. In section \ref{sectionconclusion} we present our conclusions. 
\section{Computing optimal exercise boundary
algorithm}\label{sectionibanezzapatero} 
\subsection{Introduction}
In \cite{ibanez2004mcv}, the authors propose an option pricing method
that builds a full exercise boundary as a polynomial curve whose
dimension depends on the number of underlying assets. This algorithm
consists of two phases. In the first phase the exercise boundary is
parameterized. For parameterization, the algorithm uses linear
interpolation or regression of a quadratic or cubic function at a
given exercise opportunity. In the second phase, the option price is
computed using MC simulations. These simulations are run until the
price trajectory reaches the dynamic boundary computed in the first
phase. The main advantage of this method is that it provides a full
parameterization of the exercise boundary and the exercise rule. For
American options, a buyer is mainly concerned in these values as he
can decide at ease whether or not to exercise the option. At each
exercise date $t$, the optimal exercise point $S_t^*$ is defined by
the following implicit condition, 
\begin{equation}\label{matched_condition}
P_t(S_t^{*}) = I(S_t^{*})
\end{equation}
where $P_t(x)$ is the price of the American option on the period
$[t,T]$, $I(x)$ is the exercise value (intrinsic value) of the option
and $x$ is the asset value at opportunity date $t$. As explained in
\cite{ibanez2004mcv}, these optimal values stem from the monotonicity
and convexity of the price function $P(\cdot)$ in
\refeq{matched_condition}. These are general properties satisfied by
most of the derivative securities such as maximum, minimum or
geometric average basket options. However, for the problems where the
monotonicity and convexity of the price function can not be easily
established, this algorithm has to be revisited. In the following
section we briefly discuss the sequential algorithm followed by a
proposed parallel solution. 
\subsection{Sequential Boundary Computation}\label{subsectionibanezzapaterosequential}
\begin{figure}[htb]
	\centerline{\epsfig{figure=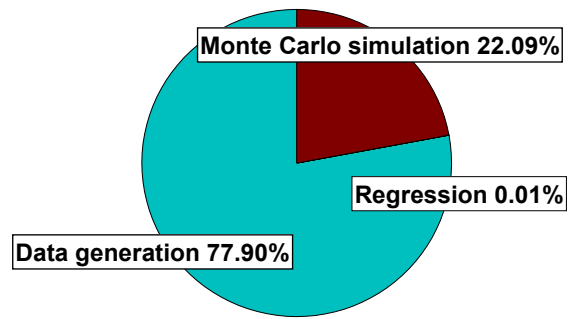,height=6cm,width=8cm}}
	\caption{The time distribution for the sequential optimal exercise boundary computation algorithm. The total time is about 40 minutes.}\label{fig:timeDistribution_OEB}
\end{figure}

The algorithm proposed in \cite{ibanez2004mcv} is used to compute the
exercise boundary. To illustrate this approach, we consider a call BA
option on the maximum of $d$ assets modeled by Geometric Brownian
Motion (GBM). It is a standard example for the multi--dimensional BA
option with maturity date $T$, constant interest rate $r$ and the
price of this option at $t_0$ is given as 
\begin{center}
$P_{t_0} = \Ee \left ( \exp{(-r\tau)} \Phi(S_{\tau},\tau) | S_{t_0} \right ) $
\end{center}
where $\tau$ is the optimal stopping time $\in \lbrace t_1,..,T
\rbrace $, defined as the first time $t_i$ such that the underlying
value $S_{\tau}$ surpasses the optimal exercise boundary at the
opportunity $\tau$ otherwise the option is held until $\tau =
\infty$. The payoff at time $\tau$ is defined as follows:
$\Phi(S_{\tau},\tau) = (\max_i(S^i_{\tau}) - K)^+$, where $i$ =
1,..,$d$, $S$ is the underlying asset price vector and $K$ is the
strike price. The parameter $d$ has a strong impact on the complexity
of the algorithm, except in some cases as the average options where
the number of dimensions $d$ can be easily reduced to one. For an
option on the maximum of $d$ assets there are $d$ separate exercise
regions which are characterized by $d$ boundaries
\cite{ibanez2004mcv}. These boundaries are monotonic and smooth curves
in $\Rr^{d-1}$. The algorithm uses backward recursive time
programming, with a finite number of exercise opportunities $m =
{1,...,N_T}$. Each boundary is computed by regression on $J$ numbers
of boundary points in $\Rr^{d-1}$. At each given opportunity, these
optimal boundary points are computed using $N_1$ MC
simulations. Further in case of an option on the maximum of $d$
underlying assets, for each asset the boundary points are computed. It
takes $n$ iterations to converge each individual point. The complexity
of this step is $\Oo \left ( \sum_{m=1}^{N_T} d \times J \times m
\times N_1 \times (N_T - m) \times n \right )$. After estimating $J$
optimal boundary points for each asset, $d$ regressions are performed
over these points to get $d$ execution boundaries. Let us assume that
the complexity of this step is $\Oo( N_T \times  regression(J))$,
where the complexity of the regression is assumed to be
constant. After computing the boundaries at all $m$ exercise
opportunities, in the second phase, the price of the option is
computed using a standard MC simulation of $N$ paths in $\Rr^d$. The
complexity of the pricing phase is $\Oo(d \times N_T \times N)$. Thus
the total complexity of the algorithm is as follows,  
\begin{flushleft}\label{equa-complexity-ibanez}
$\Oo \left ( \sum_{m=1}^{N_T} d \times J \times m \times N_1 \times (N_T - m) \times n + N_T \times  regression(J) + d \times N_T \times N \right )$\\
$\approx \Oo \left ( N_T^2 \times J \times d \times N_1 \times n + N_T \times ( J + d \times N ) \right )$
\end{flushleft}
For the performance benchmarks, we use the same simulation parameters
as given in \cite{ibanez2004mcv}. Consider a call BA option on the
maximum of $d$ assets with the following configuration. 
\begin{eqnarray}\label{ibanez}
\begin{array}{l}
\mbox {$K=100$, \textit{interest rate r = 0.05}, \textit{volatility rate $\sigma$ = 0.2},}\\
\mbox {\textit{dividend $\delta$ = 0.1}, $J=128$, $N_1=5e3$, $N=1e6$, $d=3$, }\\
\mbox {$N_T = 9$ and $T = 3$ years.}
\end{array}
\end{eqnarray}
The sequential pricing of this option \refeq{ibanez} takes 40
minutes. The distribution of the total time for the different phases
is shown in Figure \ref{fig:timeDistribution_OEB}. As can be seen, the
data generation phase, which simulates and calculates $J$ optimal
boundary points, consumes most of the computational time. We believe
that a parallel approach to this and other phases could dramatically
reduce the computational time. This inspires us to investigate a
parallel approach for I\&Z Algorithm which is presented in the
following section. The numerical results that we shall provide
indicate that the proposed parallel solution is more efficient
compared with the serial algorithm.  

\subsection{Parallel approach}\label{subsectionibanezzapateroparallel}
In this section, a simple parallel approach for I\&Z Algorithm is
presented and the pseudocode for the same is given in Algorithm
\ref{fig:flowfigure_OEB}. This approach is inspired from the solution
proposed by Muni Toke \cite{toke2006mcv}, though he presents it in the
context of a low--order homogeneous parallel cluster. The algorithm
consists of two phases. The first parallel phase is based on the
following observation: for each of the $d$ boundaries, the computation
of $J$ optimal boundary points at a given exercise date can be
simulated independently. 
\begin{algorithm}
\caption{Parallel Ibanez and Zapatero Algorithm}
\begin{algorithmic}[1]
\STATE \textbf{[glp]} Generation of the $J$ \textit{``Good Lattice Points''}
\FOR {$t=t_{N_T}$ to $t_1$}
	\FOR {$d_i=1$ to $d$} 
		\FOR {$j=1$ to $J$ \textbf{in parallel}}
			\STATE \textbf{[calc]} Computation of a boundary point with $N_1$ Monte Carlo simulations
		\ENDFOR
	\STATE \textbf{[reg]} Regression of the exercise boundary .
	\ENDFOR
\ENDFOR
\FOR{$i=1$ to $N$ \textbf{in parallel} }
	\STATE \textbf{[mc]} Computation of the partial option price. 
\ENDFOR
\STATE Estimation of the final option price by merging the partial prices.
\label{fig:flowfigure_OEB}
\end{algorithmic}
\end{algorithm}
The optimal exercise boundaries from opportunity date $m$ back to
$m-1$ are computed as follows. Note that at $m = N_T$, the boundary is
known (e.g. for a call option the boundary at $N_T$ is defined as the
strike value). Backward to $m=N_T-1$, we have to estimate $J$ optimal
points from $J$ initial good lattice points \cite{ibanez2004mcv},
\cite{haber1983pip} to regress the boundary to this time. The
regression of $\Rr^{d}$ $\rightarrow$ $\Rr^{d}$ is difficult to
achieve in a reasonable amount of time in case of large number of
training points. To decrease the size of the training set we utilize
\textit{``Good Lattice Points''} (GLPs) as described in
\cite{haber1983pip},\cite{sloan1994lmm}, and \cite{boyle:pas}. In
particular case of a call on the maximum of $d$ assets, only a
regression of $\Rr^{d-1}$ $\rightarrow$ $\Rr$ is needed, but we repeat
it $d$ times.  

The Algorithm \ref{fig:flowfigure_OEB} computes GLPs using either SSJ
library \cite{lecuyer2002ssf} or the quantification number sequences
as presented in \cite{pages2003oqq}. SSJ is a Java library for
stochastic simulation and it computes GLPs as a Quasi Monte Carlo
sequence such as Sobol or Hamilton sequences. The algorithm can also
use the number sequences readily available at
\cite{quantification}. These sequences are generated using an optimal
quadratic quantizer of the Gaussian distribution in more than one
dimension. The \textbf{[calc]} phase of the Algorithm
\ref{fig:flowfigure_OEB} is embarrassingly parallel and the $J$
boundary points are equally distributed among the computing nodes. At
each node, the algorithm simulates $N_1$ paths to compute the
approximate points. Then Newton's iterations method is used to
converge an individual approximated point to the optimal boundary
point. After computing $J$ optimal boundary points, these points are
collected by the master node, for sequential regression of the
exercise boundary. This node then repeats the same procedure for every
date $t$, in a recursive way, until $t=t_1$ in the \textbf{[reg]}
phase. 

The \textbf{[calc]} phase provides the exact optimal exercise boundary
at every opportunity date. After computation of the boundary, in the
last \textbf{[mc]} phase, the option is priced using parallel MC
simulations as shown in the Algorithm \ref{fig:flowfigure_OEB}. 

\subsection{Numerical results and performance}\label{subsection_ibanezzapatero_numericalresults}
In this section we present performance and accuracy results due to the
parallel I\&Z Algorithm described in Algorithm
\ref{fig:flowfigure_OEB}. We price a basket BA call option on the
maximum of 3 assets as given in \refeq{ibanez}. The start prices for
the assets are varied as 90, 100, and 110. The prices estimated by the
algorithm are presented in Table \ref{tab:IbanezZapa}. To validate our
results we compare the estimated prices with the prices mentioned in
Andersen and Broadies (1997) \cite{andersen2004pds}. Their results are
reproduced in the column labeled as \textit{``Binomial"}. The last
column of the table indicates the errors in the estimated prices. As
we can see, the estimated option prices are close to the desired
prices by acceptable marginal error and this error is represented by a
desirable 95\% confidence interval. 

As mentioned earlier, the algorithm relies on $J$ number of GLPs to
effectively compute the optimal boundary points. Later the
parameterized boundary is regressed over these points. For the BA
option on the maximum described in \refeq{ibanez}, Muni Toke
\cite{toke2006mcv} notes that $J$ smaller than 128 is not sufficient
and prejudices the option price. To observe the effect of the number
of optimal boundary points on the accuracy of the estimated price, the
number of GLPs is varied as shown in Table \ref{tab:impactOfJ}. For
this experiment, we set the start price of the option as $S_0 =
90$. The table indicates that increasing number of GLPs has negligible
impact on the accuracy of the estimated price. However, we observe the
linear increase in the computational time with the increase in the
number of GLPs.

[INSERT TABLE \ref{tab:IbanezZapa} HERE]

\begin{table}
\begin{center}
\begin{tabular}{|c|c|c|c|c|}
    \hline
    $S^{i}_0$ & Option Price & Variance (95\% CI) & Binomial & Error\\
	\hline
    90 & 11.254 & 153.857 (0.024) & 11.29 & 0.036 \\
	\hline
    100 & 18.378 & 192.540 (0.031) & 18.69 & 0.312 \\
    	\hline
    110 & 27.512 & 226.332 (0.035) & 27.58 & 0.068\\
	\hline
\end{tabular}
\caption{Price of the call BA on the maximum of three assets ($d=3$,
with the spot price $S^{i}_0$ for $i= 1,..,3$) using I\&Z
Algorithm. ($r=0.05$, $\delta=0.1$, $\sigma=0.2$, $\rho=0.0$, $T=3$
and $N=9$). The binomial values are referred as the true values.} 
\label{tab:IbanezZapa}
\end{center}
\end{table}

\begin{figure}[htb]
	\centerline{\epsfig{figure=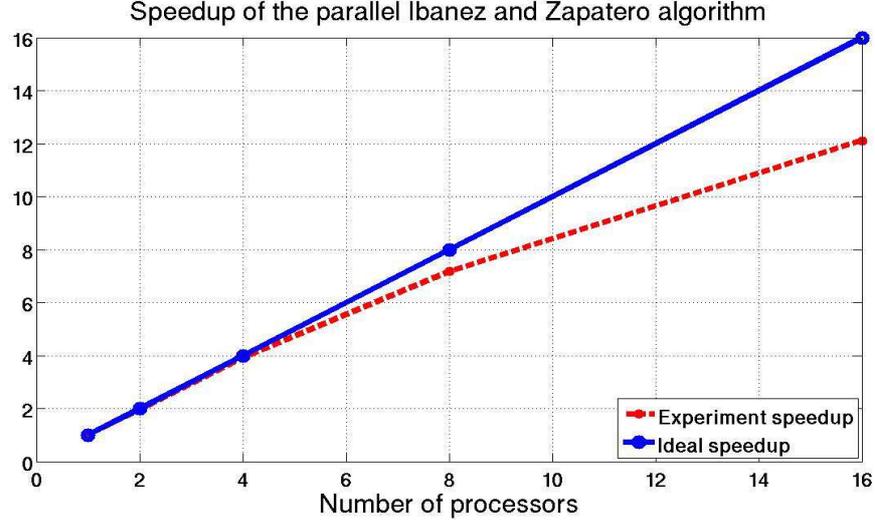,height=7cm,width=14cm}}
	\caption{Speedup of the parallel I\&Z Algorithm.}\label{fig:speedupIB}
\end{figure}

\begin{table}
\begin{center}
\begin{tabular}{|c|c|c|c|c|}
    \hline
    \textit{J} & Price & Time (in minute) & Error \\
	\hline
    128 & 11.254 & 4.6 & 0.036\\
	\hline
    256 & 11.258 & 8.1 & 0.032 \\
    	\hline
    1024 & 11.263 & 29.5 & 0.027\\
    \hline
\end{tabular}
\caption{Impact of the value of \textit{J} on the results of the
maximum on three assets option ($S_0 = 90$). The binomial price is
11.29. Running time on 16 processors.} 
\label{tab:impactOfJ}
\end{center}
\end{table}
To evaluate the accuracy of the computed prices by the parallel
algorithm, we obtained the numerical results with 16
processors. First, let us observe the effect of $N_1$, the number of
simulations required in the first phase of the algorithm, on the
computed option price. In \cite{toke2006mcv}, the author comments that
the large values of $N_1$ do not affect the accuracy of option
price. For these experiments, we set the number of GLPs, $J$, as 128
and vary $N_1$ as shown in Table \ref{tab:impactOfN1}. We can clearly
observe that $N_1$ in fact has a strong impact on the accuracy of the
computed option prices: the computational error decreases with the
increased $N_1$. A large value of $N_1$ results in more accurate
boundary points, hence more accurate exercise boundary. Further, if
the exercise boundary is accurately computed, the resulting option
prices are much closer to the true price. However this, as we can see
in the third column, comes at a cost of increased computational time. 
\begin{table}
\begin{center}
\begin{tabular}{|c|c|c|c|c|}
    \hline
    $N_1$ & Price & Time (in minute) & Error\\
	\hline
    5000 & 11.254 & 4.6 &  0.036\\	
	\hline
    10000 & 11.262 & 6.9 & 0.028\\
	\hline
    100000 & 11.276 & 35.7 & 0.014 \\
    \hline
\end{tabular}
\caption{Impact of the value of \textit{$N_1$} on the results of the
maximum on three assets option ($S_0=90$). The binomial price is
11.29. Running time on 16 processors.} 
\label{tab:impactOfN1}
\end{center}
\end{table}
The I\&Z algorithm highly relies on the accuracy and the convergence
rate of the optimal boundary points. While the former affects the
accuracy of the option price, the later affects the speed up of the
algorithm. In each iteration, to converge to the optimal boundary
point, the algorithm starts with an arbitrary point with the strike
price $K$ often as its initial value. The algorithm then uses $N_1$
random MC paths to simulate the approximated point. A convergence
criterion is used to optimize this approximated point. The simulated
point is assumed to be optimal when it satisfies the following
condition, $\vert S_{t_n}^{i,(simulated)} - S_{t_n}^{i,(initial)}
\vert < \epsilon$ = 0.01, where the $S_{t_n}^{i,(initial)}$ is the
initial point at a given opportunity $t_n$, $i=1..J$ and the
$S_{t_n}^{i,(simulated)}$ is the point simulated by using $N_1$ MC
simulations. In case, the condition is not satisfied, this procedure
is repeated and now with the initial point as the newly simulated
point $S_{t_n}^{i,(simulated)}$. Note that the number of iterations
$n$, required to reach to the optimal value, varies depending on the
fixed precision in the Newton procedure (for instance, with a
precision $\epsilon = 0.01$, $n$ varies from 30 to 60). We observed
that not all boundary points take the same time for the
convergence. Some points converge faster to the optimal boundary
points while some take longer than usual. Since the algorithm has to
wait until all the points are optimized, the slower points increase
the computational time, thus reducing the efficiency of the parallel
algorithm, see Figure \ref{fig:speedupIB}. 

\section{The Classification and Monte Carlo algorithm}\label{sectionpicazo}
\subsection{Introduction}\label{subsectionpicazointro}
The Monte Carlo approaches for BA option pricing are usually based on
continuation value computation \cite{longstaff:vao} or continuation
region estimation \cite{hickernell2002mca}, \cite{ibanez2004mcv}. The
option holder decides either to execute or to continue with the
current option contract based on the computed asset value. If the
asset value is in the exercise region, he executes the option
otherwise he continues to hold the option. Denote that the asset
values which belong to the exercise region will form the exercise
values and rest will belong to the continuation region. In
\cite{hickernell2002mca} Picazo et al. propose an algorithm based on
the observation that at a given exercise opportunity the option holder
makes his decision based on whether the sign of $(exercise\ value -
continuation\ value)$ is positive or negative. The author focuses on
estimating the continuation region and the exercise region by
characterizing the exercise boundary based on these signs. The
classification algorithm is used to evaluate such sign values at each
opportunity. In this section we briefly describe the sequential
algorithm described in \cite{hickernell2002mca} and propose a parallel
approach followed by performance benchmarks.      
\subsection{Sequential algorithm}\label{subsectionpicazosequential}
For illustration let us consider a BA option on $d$ underlying assets
modeled by Geometric Brownian Motion (GBM). $S_t = (S_t^i)$ with
$i=1,..,d$. The option price at time $t_0$ is defined as follows: 
\begin{center}
$P_{t_0}(S_{t_0}) = \Ee \left (\exp{(-r\tau)} \Phi(S_{\tau},\tau) | S_{t_0} \right ) $
\end{center}
where $\tau$ is the optimal stopping time $\in \lbrace t_1,..,T
\rbrace $, $T$ is the maturity date, $r$ is the constant interest rate
and $\Phi(S_{\tau},\tau)$ is the payoff value at time $\tau$. In case
of I\&Z Algorithm, the optimal stopping time is defined when the
underlying asset value crosses the exercise boundary. The CMC
algorithm defines the stopping time whenever the underlying asset
value makes the sign of $(exercise\ value - continuation\ value)$
positive. Without loss of generality, at a given time $t$ the BA
option price on the period $[t,T]$ is given by: 
\begin{center}
$P_{t}(S_t) = \Ee \left (\exp{(-r(\tau - t))} \Phi(S_{\tau},\tau) | S_{t} \right ) $
\end{center}
where $\tau$ is the optimal stopping time $\in \lbrace 1,..,T \rbrace
$. Let us define the difference between the payoff value and the
option price at time $t_m$ as, 
\begin{center}
$\beta(t_m, S_{t_m}) = \Phi(S_{t_m},t_m) - P_{t_m}(S_{t_m}) $ 
\end{center}
where $m \in \lbrace 1,..,T \rbrace$. The option is exercised when
$S_{t_m} \in \lbrace x | \beta(t_m, x) > 0 \rbrace$ which is the
exercise region, and $x$ is the simulated underlying asset value,
otherwise the option is continued. The goal of the algorithm is to
determinate the function $\beta(\cdot)$ for every opportunity
date. However, we do not need to fully parameterize this function. It
is enough to find a function $F_t(\cdot)$ such that $sign F_t(\cdot) =
sign \beta(t, \cdot)$. 

The algorithm consists of two phases. In the first phase, it aims to
find a function $F_t(\cdot)$ having the same sign as the function
$\beta(t, \cdot)$. The AdaBoost or LogitBoost algorithm is used to
characterize these functions. In the second phase the option is priced
by a standard MC simulation by taking the advantage of the
characterization of $F_{t_m}(\cdot)$, so for the $(i)^{th}$ MC
simulation we get the optimal stopping time $\tau_{(i)} = \min \lbrace
t_m \in \lbrace t_1,t_2,...,T \rbrace | F_t(S^{(i)}_t) > 0
\rbrace$. The $(i)$ is not the index of the number of assets.   

Now, consider a call BA option on the maximum of $d$ underlying assets
where the payoff at time $\tau$ is defined as $\Phi(S_{\tau},\tau) =
(\max_i(S^i_{\tau}) - K)^+$ with $i=1,..,d$. During the first phase of
the algorithm, at a given opportunity date $t_m$ with $m \in
{1,...,N_T}$, $N_1$ underlying price vectors sized $d$ are
simulated. The simulations are performed recursively in backward from
$m = T$ to $m = 1$. From each price point, another $N_2$ paths are
simulated from a given opportunity date to the maturity date to
compute the ``small'' BA option price at this opportunity
(i.e. $P_{t_m}(S_{t_m})$). At this step, $N_1$ option prices related
to the opportunity date are computed. The time step complexity of this
step is $\Oo(N_1 \times d \times m \times N_2 \times (N_T - m))$. In
the classification phase, we use a training set of $N_1$ underlying
price points and their corresponding option prices at a given
opportunity date. In this step, a non--parametric regression is done
on $N_1$ points to characterize the exercise boundary. This first
phase is repeated for each opportunity date. In the second phase, the
option value is computed by simulating a large number, $N$, of
standard MC simulations with $N_T$ exercise opportunities. The
complexity of this phase is $\Oo(d \times N_T \times N)$. Thus, the
total time steps required for the algorithm can be given by the
following formula, 
\begin{flushleft}\label{equa-complexity-picazo}
$\Oo \left ( \sum_{m=1}^{N_T} N_1 \times d \times m \times N_2 \times (N_T - m) + N_T \times classification(N_1) + d \times N_T \times N \right )$\\
$\approx \Oo \left ( N_T^2 \times N_1 \times d \times N_2 + N_T \times ( N_1 + d \times N ) \right )$
\end{flushleft}
where $\Oo(classification(\cdot))$ is the complexity of the classification phase and the details of which can be found in \cite{hickernell2002mca}. For the simulations, we use the same option parameters as described in \refeq{ibanez}, taken from \cite{ibanez2004mcv}, and the parameters for the classification can be found in \cite{hickernell2002mca}.
\begin{eqnarray}\label{picazo}
\begin{array}{l}
\mbox {$K=100$, \textit{interest rate r = 0.05}, \textit{volatility rate $\sigma$ = 0.2},}\\
\mbox {\textit{dividend $\delta$ = 0.1}, $N_1=5e3$, $N_2=500$, $N=1e6$, $d=3$ }\\
\mbox {$N_T = 9$ and $T = 3$ years.}
\end{array}
\end{eqnarray}
Each of the $N_1$ points of the training set acts as a seed which is
further used to simulate $N_2$ simulation paths. From the exercise
opportunity $m$ backward to $m-1$, a Brownian motion bridge is used to
simulate the price of the underlying asset. The time distribution of
each phase of the sequential algorithm for pricing the option
\refeq{picazo} is shown in Figure
\ref{fig:timeDistributionfigure_PI}. As we can see from the figure,
the most computationally intensive part is the data generation phase
which is used to compute the option prices required for
classification. In the following section we present a parallel
approach for this and rest of the phases of the algorithm.  
\subsection{Parallel approach}\label{subsectionpicazoparallel}
The Algorithm \ref{fig:flowfigure_CMC} illustrates the parallel
approach based on CMC Algorithm. At $t_m = T$ we generate $N_1$ points
of the price of the underlying assets, $S_{t_m}^{(i)}$, $i=1,..,N_1$
then apply the Brownian bridge simulation process to get the price at
the backward date, $t_{m-1}$. For simplicity we assume a
master--worker programming model for the parallel implementation: the
master is responsible for allocating independent tasks to workers and
for collecting the results. The master divides $N_1$ simulations into
$nb$ tasks then distributes them to a number of workers. Thus each
worker has $N_1/nb$ points to simulate in the \textbf{[calc]}
phase. Each worker, further, simulates $N_2$ paths for each point from
$t_m$ to $t_{N_T}$ starting at $S_{t_m}^{(i)}$ to compute the option
price related to the opportunity date. Next the worker calculates the
value $y_j = (exercise\ value - continuation\ value)$,
$j=1,..,N_1/nb$. The master collects the $y_j$ of these $nb$ tasks
from the workers and then classifies them in order to return the
characterization model of the associated exercise boundary in the
\textbf{[class]} phase. 
\begin{algorithm}
\caption{Parallel Classification and Monte Carlo Algorithm}
\begin{algorithmic}[1]
\FOR {$t=t_{N_T}$ to $t_1$}
	\FOR {$i=1$ to $N_1$ \textbf{in parallel}}
		\STATE \textbf{[calc]} Computation of training points. 
	\ENDFOR
	\STATE \textbf{[class]} Classification using boosting.
\ENDFOR
\FOR{$i=1$ to $N$ \textbf{in parallel} }
	\STATE \textbf{[mc]} The partial option price computation.
\ENDFOR
\STATE Estimation of the final option price by merging the partial prices.
\label{fig:flowfigure_CMC}
\end{algorithmic}
\end{algorithm}
For the classification phase, the master does a non-parametric
regression with the set $(x_{(i)},y_{(i)})^{N_1}_{i=1}$, where
$x_{(i)} = S_{t_m}^{(i)}$,  to get the function $F_{t_m}(x)$ described
above in Section \refeq{subsectionpicazosequential}. The algorithm
recursively repeats the same procedure for earlier time intervals
$[m-1,1]$. As a result we obtain the characterization of the
boundaries, $F_{t_m}(x)$, at every opportunity $t_m$. Using these
boundaries, a standard MC simulation, \textbf{[mc]}, is used to
estimate the option price. The MC simulations are distributed among
workers such that each worker has the entire characterization boundary
information $(F_{t_m}(x), m=1,..,N_T)$ to compute the partial option
price. The master later merges the partially computed prices and
estimates the final option price. 
\subsection{Numerical results and performance}\label{subsection_picazo_performance}
\begin{figure}[htb]
	\centerline{\epsfig{figure=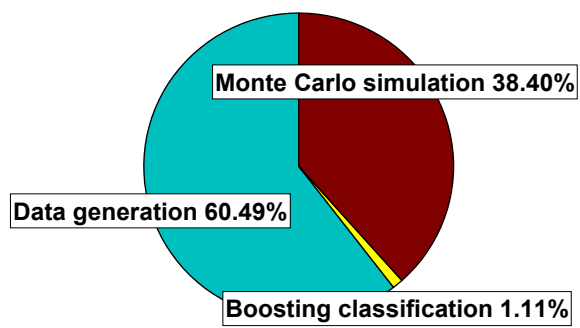,height=6cm,width=8cm}}
	\caption{The time distribution for different phases of the sequential Classification--Monte Carlo algorithm. The total time is about 30 minutes.}\label{fig:timeDistributionfigure_PI}
\end{figure}

\begin{figure}[htb]
	\centerline{\epsfig{figure=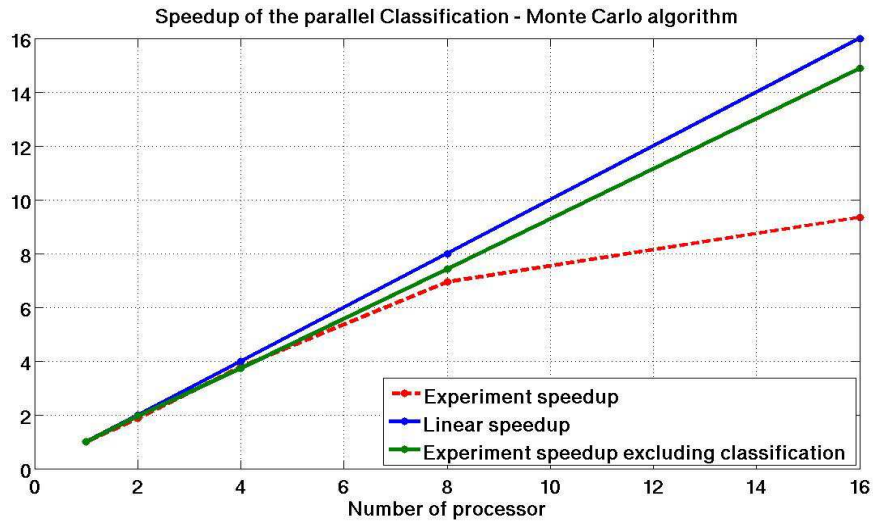,height=7cm,width=14cm}}
	\caption{Speedup of the parallel CMC Algorithm.}\label{fig:speedupPicazo}
\end{figure}

In this section we present the numerical and performance results of
the parallel CMC Algorithm. We focus on the standard example of a call
option on the maximum of 3 assets as given in \refeq{picazo}. As it
can be seen, the estimated prices are equivalent to the reference
prices presented in Andersen and Broadies \cite{andersen2004pds},
which are represented in the \textit{``Binomial''} column in Table
\ref{tab:cmcmax}. For pricing this option, the sequential execution
takes up to 30 minutes and the time distribution for the different
phases can be seen in Figure \ref{fig:timeDistributionfigure_PI}. The
speed up achieved by the parallel algorithm is presented in Figure
\ref{fig:speedupPicazo}. We can observe from the figure that the
parallel algorithm achieves linear scalability with a fewer number of
processors. The different phases of the algorithm scale
differently. The MC phase being embarrassingly parallel scales
linearly, while, the number of processors has no impact on the
scalability of the classification phase. The classification phase is
sequential and takes a constant amount of time for the same
option. This affects the overall speedup of the algorithm as shown in
Figure \ref{fig:speedupPicazo}.   
\begin{table}
\begin{center}
\begin{tabular}{|c|c|c|c|c|}
    \hline
    $S_0$ & Price & Variance (95\% CI) & Binomial & Error\\
	\hline
    90 & 11.295 & 190.786 (0.027) & 11.290 & 0.005\\
	\hline
    100 & 18.706 & 286.679 (0.033) & 18.690 & 0.016\\
    	\hline
    110 & 27.604 & 378.713 (0.038) & 27.580 & 0.024 \\
	\hline
\end{tabular}
\caption{Price of the call BA on the maximum of three assets using CMC Algorithm. ($r=0.05$, $\delta=0.1$, $\sigma=0.2$, $\rho=0.0$, $T=3$, $N=9$ opportunities)}
\label{tab:cmcmax}
\end{center}
\end{table}

\section{Conclusion}\label{sectionconclusion}
The aim of the study is to develop and implement parallel Monte Carlo
based Bermudan/American option pricing algorithms. In this paper, we
particularly focused on multi--dimensional options. We evaluated the
scalability of the proposed parallel algorithms in a computational
grid environment. We also analyzed the performance and the accuracy of
both algorithms. While I\&Z Algorithm computes the exact exercise
boundary, CMC Algorithm estimates the characterization of the
boundary. The results obtained clearly indicate that the scalability
of I\&Z Algorithm is limited by the boundary points computation. The
Table \ref{tab:impactOfJ} showed that there is no effective advantage
to increase the number of such points as will, just to take advantage
of a greater number of available CPUs. Moreover, the time required for
computing individual boundary points varies and the points with slower
convergence rate often haul the performance of the algorithm. However,
in the case of CMC Algorithm, the sequential classification phase
tends to dominate the total parallel computational time. Nevertheless,
CMC Algorithm can be used for pricing different option types such as
maximum, minimum or geometric average basket options using a generic
classification configuration. While the optimal exercise boundary
structure in I\&Z Algorithm needs to be tailored as per the option
type and requires. Parallelizing classification phase presents us
several challenges due to its dependency on inherently sequential
non--parametric regression. Hence, we direct our future research to
investigate efficient parallel algorithms for computing exercise
boundary points, in case of I\&Z Algorithm, and the classification
phase, in case of CMC Algorithm. 
\section{Acknowledgments}
This research is supported by the French ``ANR-CIGC GCPMF'' project and Grid5000 has been funded by ACI-GRID.

\end{document}